# WHAT WE CAN LEARN FROM DNS OF TURBULENCE IN POROUS MEDIA: MODELING TURBULENT FLOW IN COMPOSITE POROUS/FLUID DOMAINS

Kuznetsov A.V.
Department of Mechanical and Aerospace Engineering,
North Carolina State University,
Campus Box 7910,
Raleigh, NC 27695-7910, USA
E-mail: avkuznet@ncsu.edu

## ABSTRACT

Our recently published papers reporting results of Direct Numerical Simulation (DNS) of forced convection flows in porous media suggest that in a porous medium the size of turbulent structures is restricted by the pore scale. Since the turbulent kinetic energy is predominantly contained within large eddies, this suggests that turbulent flow in a porous medium may carry less energy that its counterpart in a clear fluid domain.

We use this insight to develop a practical model of turbulent flow in composite porous/fluid domains. In such domains, most of the flow is expected to occur in the clear fluid region; therefore, in most cases the flow in the porous region either remains laminar or starts its transition to turbulence even if the flow in the clear fluid region is fully turbulent. This conclusion is confirmed by comparing appropriate Reynolds numbers with their critical values. Therefore, for most cases, using the Forchheimer term in the momentum equation and the thermal dispersion term in the energy equation may result in a sufficiently good model for the porous region. However, what may really affect turbulent convection in composite domains is the roughness of the porous/fluid interface. If particles or fibers that constitute the porous medium (and the pores) are relatively large, the impact of the roughness on convection heat transfer in composite porous/fluid domains may be much more significant than the impact of possible turbulence in the porous region. We use the above considerations to develop a practical model of turbulent flow in a composite porous/fluid domain, concentrating on the effect of interface roughness on turbulence.

## INTRODUCTION

Turbulence in porous media is an extensively studied subject, please see a recently published book on convection in porous media [1]. The book by de Lemos [2] specifically addresses convection in porous media. This topic was also addressed in many review chapters [3-6]. The importance of this topic is related to its relevance to many applications, such as forestry (forest fires), chemical reactor design, agricultural engineering, catalytic converters, bio-filters, crude oil extraction, biomechanics of porous organs, and many others [7].

The first physically-based turbulence model was developed in a classical paper by Antohe and Lage [8]. This paper started the field of turbulence modeling in porous media. Currently, there are two approaches to modeling turbulence in porous media. According to the first approach, expressed in [9, 10], true macroscopic turbulence in a dense porous medium is impossible because of the limitation on the size of turbulent eddies imposed by the pore scale. This limitation prevents the transfer of turbulent kinetic energy from larger to smaller turbulent eddies. For this reason, one can only talk about turbulence within the pores.

The second approach simulates macroscopic turbulence. Representative models of this type are those developed in [8, 11-13]. Turbulence models of this type (macroscopic models) have also been used to simulate flow in a porous matrix represented by a periodic array of square cylinders [14, 15]. A similar approach was developed in [16], where a large-scale model of turbulence in porous media was developed utilizing the renormalization group method.

It should also be noted that a large number of turbulence models combine the above two approaches, simulating both pore-scale turbulence and large-scale turbulence. Representative models are those developed by M. de Lemos and colleagues [17-34].

We recently investigated whether macroscopic turbulence is possible in a porous medium by conducting a series of DNS studies of turbulent flow in a specially designed porous matrix [35-37]. The advantage of our approach is that DNS avoids any kind of turbulence modeling. The results that we obtained suggest that the size of turbulent structures is restricted by the pore scale. We used three methods to prove this result, utilizing three independent techniques of analyzing turbulent length scales (two-point correlations, integral length scales, and pre-multiplied energy spectra). Two different DNS methods, which complement and verify each other, were utilized in [35-37]: (i) the finite volume method (FVM); this methods directly solves the Navier-Stokes equations and (ii) the Lattice-Boltzmann method (LBM); this method indirectly corresponds to solving the Navier-Stokes equations. Both DNS methods have second order accuracy in space and time.

We also investigated what implications our findings have on turbulence modeling. Our preliminary results suggest that the effect of turbulence in porous media can be approximately modeled by the Forchheimer term [38]. This is in line with the model proposed by Masuoka and Takatsu [39].

There is also considerable interest in turbulent flows in composite porous/fluid domains. In such domains, most of the flow is expected to occur in the clear fluid region. In most cases the flow in the porous region either remains laminar or starts its transition to turbulence even if the flow in the clear fluid region is fully turbulent. Based on our DNS results, we propose that using the Brinkman-Forchheimer equation for modeling flow in a porous region is an acceptable approximation. Thus our approach is to use the turbulent model in the clear fluid region and the Brinkman-Forchheimer model in the porous region, and then match the solutions at the porous/fluid interface.

We developed this approach in [40-47]. Our purpose here is to review this approach, concentrating on modeling the effect of roughness of the porous/fluid interface. If particles or fibers that constitute the porous medium are large, the impact of interface roughness on convection heat transfer in composite porous/fluid domains may be more significant than the impact of possible turbulence in the porous region.

## NUMERICAL METHOD

In Fig. 1 we show a composite circular duct whose central portion, $0 \le r \le \xi R$, is occupied by a clear fluid and peripheral potion, $\xi R \le r \le R$, is occupied by an isotropic fluid-saturated porous medium of uniform porosity. The wall of the duct is subject to either uniform heat flux or uniform wall temperature. We divided the flow domain into two regions. The first region is the central region, where the flow is turbulent. The second region is the peripheral porous region, where the flow is assumed to be laminar or, even if it is turbulent, to be adequately simulated by the Brinkman-Forchheimer equation. We assume hydrodynamically and thermally fully developed flow. We closely follow the approach that we developed in [42].

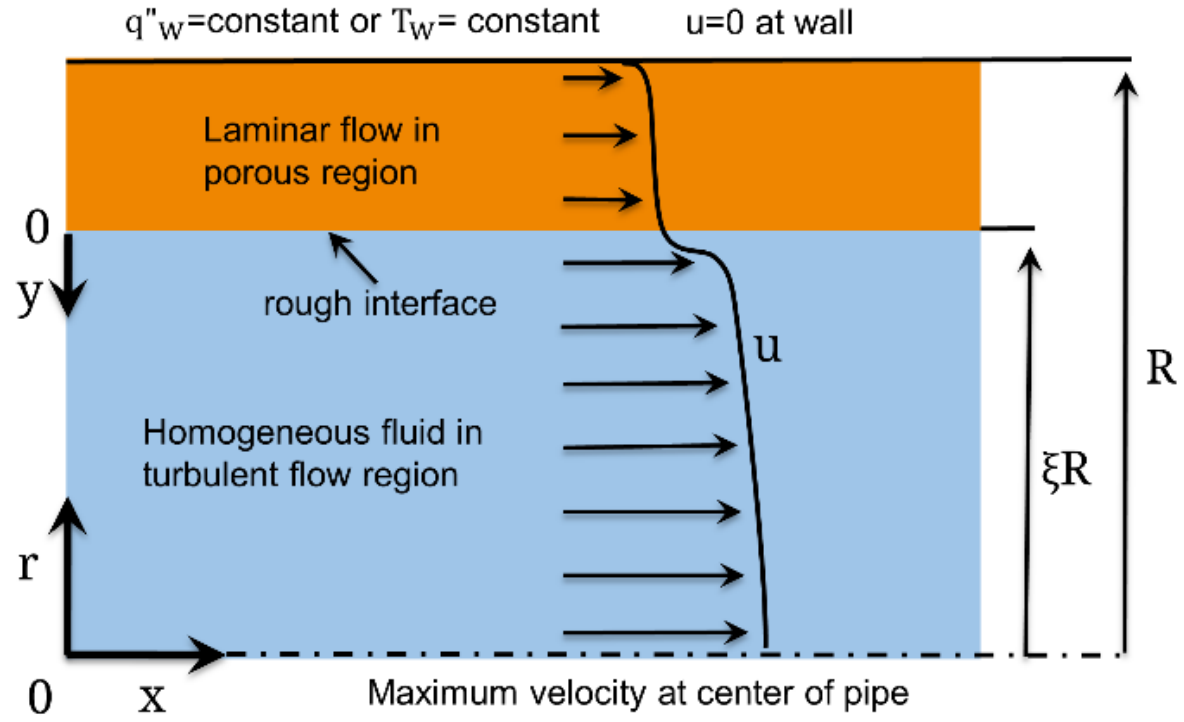

**Figure 1** Schematic diagram of the problem.

### Modeling Momentum Transport in the Clear Fluid Region

Turbulent flow in the clear fluid region can be computed by using the following equation [48]:

$$\frac{du^+}{dy^+} = \frac{1}{1+\nu_T^+}\left(1 - \frac{y^+}{\xi R^+}\right) \tag{1}$$

Here $u^+$ is the dimensionless velocity, $u/u_\tau$; $u$ is the longitudinal velocity; $u_\tau$ is the friction velocity at the porous/fluid interface, $\sqrt{\tau_i/\rho_f}$; $\tau_i$ is the shear stress at the porous/fluid interface (at $r = \xi R$); $\rho_f$ is the fluid density; $R^+$ is the dimensionless radius of the duct, $u_\tau R/\nu_f$; $R$ is the duct radius; $\nu_f$ is the fluid kinematic viscosity; $y^+$ is the dimensionless distance from the porous/fluid interface towards the duct center, $\xi R^+ - r^+$; $r^+$ is the dimensionless radial coordinate, $u_\tau r/\nu_f$; $r$ is the radial coordinate; $\nu_T^+$ is the dimensionless eddy viscosity, $\nu_T/\nu_f$; $\nu_f$ is the fluid kinematic viscosity; and $\nu_T$ is the eddy diffusivity of momentum.

We followed the proposal by Durbin et al. [49] and used a combination of *k-l* (in the vicinity of the interface) and $k-\varepsilon$ (in the rest of the domain) models to account for the roughness of the interface.

In the central part of the clear fluid region, $y^+_{match} \le y^+ \le R^+$ ($y^+_{match}$ is defined later on), we utilized the following $k-\varepsilon$ model

$$\nu_T^+\left(\frac{\partial u^+}{\partial r^+}\right)^2 - \varepsilon^+ + \frac{1}{r^+}\frac{\partial}{\partial r^+}\left[r^+\left(1+\nu_T^+\right)\frac{\partial k^+}{\partial r^+}\right] = 0 \tag{2}$$

where

$$k^+ = k/u_\tau^2,\ \varepsilon^+ = \varepsilon\nu_f/u_\tau^4 \tag{3}$$

We used the following equation to model the dissipation rate equation of turbulence kinetic energy:

$$C_{\varepsilon 1}\frac{\varepsilon^+}{k^+}\nu_T^+\left(\frac{\partial u^+}{\partial r^+}\right)^2 - C_{\varepsilon 2}\frac{\left(\varepsilon^+\right)^2}{k^+} + \frac{1}{r^+}\frac{\partial}{\partial r^+}\left[r^+\left(1+\frac{\nu_T^+}{\sigma_\varepsilon}\right)\frac{\partial \varepsilon^+}{\partial r^+}\right] = 0 \tag{4}$$

We found the dimensionless eddy viscosity from the following equation:

$$\nu_T^+ = C_\mu\frac{\left(k^+\right)^2}{\varepsilon^+} \tag{5}$$

We used the following values of modeling constants:

$$C_{\varepsilon 1} = 1.44,\ C_{\varepsilon 2} = 1.92,\ \sigma_\varepsilon = 1.3,\ C_\mu = 0.09 \tag{6}$$

In the region close to the fluid/porous interface (on the fluid side of the interface), we used the following *k-l* model:

$$\varepsilon^+ = \frac{\left(k^+\right)^{3/2}}{l_\varepsilon^+} \tag{7}$$

where

$$l_\varepsilon^+ = l_\varepsilon\frac{u_\tau}{\nu} \tag{8}$$

Eq. (7) replaces Eq. (4) for $0 \le y^+ \le y^+_{match}$, and Eq. (5) is replaced with the following equation:

$$\nu_T^+ = C_\mu \left(k^+\right)^{1/2} l_\nu^+ \quad (9)$$

where

$$l_\nu^+ = l_\nu \frac{u_\tau}{\nu} \quad (10)$$

We used the Van Driest form of length scales:

$$l_\varepsilon^+ = C_l y_{eff}^+ \left(1 - e^{-R_y/A_\varepsilon}\right) \quad (11)$$

and

$$l_\nu^+ = C_l y_{eff}^+ \left(1 - e^{-R_y/A_\nu}\right) \quad (12)$$

where

$$y_{eff}^+ = y_{eff} \frac{u_\tau}{\nu_f} \quad (13)$$

and

$$R_y = y_{eff} k^{1/2} / \nu_f = y_{eff}^+ \left(k^+\right)^{1/2} \quad (14)$$

The parameter $y_{eff}^+$ in Eqs. (11)-(14) is the modified dimensionless distance from the interface:

$$y_{eff}^+ = y^+ + y_0^+ \quad (15)$$

Here $y_0^+$ is the dimensionless hydrodynamic roughness. We related this parameter to the dimensionless equivalent sand-grain roughness parameter, $k_s^+$ ($= k_s u_\tau / \nu_f$), by the equation given in [50]:

$$y_0^+ = 0 \ \text{ if } \ k_s^+ \le 4.535 \quad (16a)$$

$$y_0^+ = 0.9\left[\sqrt{k_s^+} - k_s^+ \exp\left(-k_s^+/6\right)\right] \quad \text{if} \ \ k_s^+ > 4.535 \quad (16b)$$

The average diameter of a solid particle that constitutes the porous medium was estimated by using the Carman-Kozeny equation [1]:

$$d_p = \frac{\sqrt{180}\left(1-\phi\right) K^{1/2}}{\phi^{3/2}} \quad (17)$$

where $K$ is the permeability and $\phi$ is the porosity.

Assuming that the equivalent sand-grain roughness parameter, $k_s$, can be estimated as $d_p/2$, the following equation for $k_s^+$ is obtained:

$$k_s^+ = \frac{u_\tau}{\nu} k_s = \frac{u_\tau}{\nu} \frac{\sqrt{180}\left(1-\phi\right)}{2\phi^{3/2}} K^{1/2} = R^+ \frac{6.70\left(1-\phi\right)}{\phi^{3/2}} \mathrm{Da}^{1/2} \quad (18)$$

Here Da is the Darcy number, $K/R^2$.

The modeling constants in the *k-l* model are

$$C_l = 2.5,\ A_\varepsilon = 5.0,\ A_\nu^0 = 62.5,\ A_\nu = \max\left[1, A_\nu^0\left(1 - k_s^+/90\right)\right] \quad (19)$$

The dimensionless distance from the interface where the *k-l* model must be switched to $k-\varepsilon$ model is [49]:

$$y_{match}^+ = \log\left(20\right) A_\nu / \left(k^+\right)^{1/2} \quad (20)$$

The $k$ and $\varepsilon$ equations are solved subject to the following boundary conditions:

$$k^+\left(0\right) = \frac{1}{\sqrt{C_\mu}} \min\left[1, \left(k_s^+/90\right)^2\right] \quad (21)$$

Here $k_s^+$ is given by equation (18). In the center of the pipe the following conditions are utilized:

$$\partial k^+ / \partial r^+ = 0 \ \text{ and } \ \partial \varepsilon^+ / \partial r^+ = 0 \quad (22)$$

In the matching point $k^+$ and $\varepsilon^+$ must be continuous.

**Modeling Momentum Transport in the Porous Region**

We used the Brinkman-Forchheimer equation [1] to model flow in the porous region. We used the same dimensionless variables as in the clear fluid region:

$$\frac{2}{\xi R^+} + \left(\frac{\mu_{eff}}{\mu_f}\right) \frac{1}{r^+} \frac{d}{dr^+}\left(r^+ \frac{du^+}{dr^+}\right) - \frac{u^+}{\mathrm{Da}\left(R^+\right)^2} - \frac{c_F}{\mathrm{Da}^{1/2} R^+}\left(u^+\right)^2 = 0 \quad (23)$$

Here $c_F$ is the Forchheimer coefficient, $\mu_f$ is the fluid dynamic viscosity, and $\mu_{eff}$ is the effective dynamic viscosity in the porous region.

**Modeling Thermal Energy Transport in the Clear Fluid Region**

We used a model which is based on a constant turbulent Prandtl number approximation. For the uniform wall heat flux case, the energy equation can be presented as:

$$\frac{1}{r^+} \frac{d}{dr^+}\left[\left(1 + \nu_T^+ \frac{\mathrm{Pr}}{\mathrm{Pr}_t}\right) r^+ \frac{d\theta}{dr^+}\right] = -\frac{1}{\left(R^+\right)^2} \frac{u^+}{U_m^+} \quad (24)$$

Here Pr is the Prandtl number, $\nu_f / a_f$; $a_f$ is the fluid thermal diffusivity; $\mathrm{Pr}_t$ is the turbulent Prandtl number, $\nu_T / a_T$; and $a_T$ is the eddy diffusivity of heat. The mean fluid velocity in the duct, $U_m^+$, is defined as:

$$U_m^+ = \frac{2}{\left(R^+\right)^2} \int_0^{R^+} u^+ r^+ dr^+ \quad (25)$$

In Eq. (24), $\theta$ is the dimensionless temperature for the uniform heat flux case, which is defined as:

$$\theta = (1/\mathrm{Nu})(T - T_W)/(T_m - T_W) \qquad (26)$$

Here $T$ is the temperature, $T_W$ is the wall temperature (at $r^+ = R^+$), $T_m$ is the mean temperature in the duct:

$$T_m = \frac{2}{R^2 U_m}\int_0^R uTrdr \qquad (27)$$

where Nu is the Nusselt number:

$$\mathrm{Nu} = h2R/k_f = 2Rq''/\left[k_f\left(T_W - T_m\right)\right] \qquad (28)$$

and $h$ is the heat transfer coefficient.

For the uniform wall temperature case the thermal energy equation is:

$$\frac{1}{r^+}\frac{d}{dr^+}\left[\left(1+\mu_T^+\frac{\mathrm{Pr}}{\mathrm{Pr}_t}\right)r^+\frac{d\varphi}{dr^+}\right] = -\frac{1}{\left(R^+\right)^2}\mathrm{Nu}\varphi\frac{u^+}{U_m^+} \qquad (29)$$

where

$$\varphi = (T - T_W)/(T_m - T_W) \qquad (30)$$

is the dimensionless temperature for the uniform wall temperature case.

**Modeling Thermal Energy Transport in the Porous Region**

For the uniform wall heat flux case, we used the following form the thermal energy equation in the porous region:

$$\frac{1}{r^+}\frac{d}{dr^+}\left[\left(\frac{k_m}{k_f} + C\,\mathrm{Pr}\,\mathrm{Re}_p\,u^+\right)r^+\frac{d\theta}{dr^+}\right] = -\frac{1}{\left(R^+\right)^2}\frac{u^+}{U_m^+} \qquad (31)$$

Here $C$ is the dimensionless experimental constant in the correlation for thermal dispersion; $k_f$ is the fluid thermal conductivity; $k_m$ is the stagnant thermal conductivity of the porous medium (when $u^+ = 0$); $\mathrm{Re}_p = u_\tau d_p / \nu_f$ is the Reynolds number based on the average particle diameter, $d_p$, and the friction velocity at the porous/fluid interface, $u_\tau$.

For the uniform wall temperature case we used the following form of the thermal energy equation for the porous region:

$$\frac{1}{r^+}\frac{d}{dr^+}\left[\left(\frac{k_m}{k_f} + C\,\mathrm{Pr}\,\mathrm{Re}_p\,u^+\right)r^+\frac{d\varphi}{dr^+}\right] = -\frac{1}{\left(R^+\right)^2}\mathrm{Nu}\varphi\frac{u^+}{U_m^+} \qquad (32)$$

**Compatibility Condition**

After we determined the velocity and temperature distributions, the Nusselt number was found utilizing a compatibility condition [51]. For the uniform wall heat flux case the compatibility condition is

$$\mathrm{Nu} = U_m^+\left(R^+\right)^2 / \left[2\int_0^{R^+} u^+\,\theta\,r^+dr^+\right] \qquad (33)$$

For the uniform wall temperature case the compatibility condition takes the following form:

$$\mathrm{Nu} = -2\frac{k_m}{k_f}R^+\left.\frac{d\varphi}{dr^+}\right|_{r^+ = R^+} \qquad (34)$$

**Boundary Conditions**

At the wall of the pipe, $r^+ = R^+$, we used the no-slip condition:

$$u^+ = 0 \qquad (35)$$

The definitions of the dimensionless temperatures require that for the uniform wall heat flux case at $r^+ = R^+$:

$$\theta = 0 \qquad (36)$$

and for the uniform wall temperature case at $r^+ = R^+$:

$$\varphi = 0 \qquad (37)$$

From the definition of $u^+$ and $r^+$ we obtained that

$$\left.\frac{\partial u^+}{\partial r^+}\right|_{r^+ = \xi R^+ - 0} = -\frac{1}{1+\left.\nu_T^+\right|_{r^+ = \xi R^+}} \qquad (38)$$

The jump in the shear stress condition suggested in [52, 53] leads to the following equation:

$$\left(\frac{\nu_{eff}}{\nu_f + \left.\nu_T\right|_{r^+ = \xi R^+}}\right)\left.\frac{\partial u^+}{\partial r^+}\right|_{r^+ = \xi R^+ + 0} - \left.\frac{\partial u^+}{\partial r^+}\right|_{r^+ = \xi R^+ - 0} = \frac{\beta}{\mathrm{Da}^{1/2}R^+}u_i^+ \qquad (39)$$

Here $u_i^+$ is the dimensionless filtration velocity at the interface, and $\beta$ is the dimensionless adjustable coefficient [35, 36].

The substitution of Eq. (38) into Eq. (39) gives the following equation:

$$\left(\frac{\nu_{eff}/\nu_f}{1+\left.\nu_T^+\right|_{\xi R^+}}\right)\left.\frac{\partial u^+}{\partial r^+}\right|_{\xi R^+ + 0} + \frac{1}{1+\left.\nu_T^+\right|_{\xi R^+}} = \frac{\beta}{\mathrm{Da}^{1/2}R^+}u_i^+ \qquad (40)$$

In addition to Eq. (40), we imposed the continuity of the filtration velocity, temperature, and heat flux at the interface, $r^+ = \xi R^+$:

$$\left.u^+\right|_{r^+ = \xi R^+ - 0} = \left.u^+\right|_{r^+ = \xi R^+ + 0} = u_i^+ \qquad (41)$$

For the uniform wall heat flux case we required that at the interface:

$$\theta\big|_{r^+=\xi R^+-0} = \theta\big|_{r^+=\xi R^++0}, \qquad \left.\frac{\partial\theta}{\partial r^+}\right|_{r^+=\xi R^+-0} = \frac{k_{eff}}{k_f}\left.\frac{\partial\theta}{\partial r^+}\right|_{r^+=\xi R^++0} \tag{42}$$

For the uniform wall temperature case we required that at the interface:

$$\varphi\big|_{r^+=\xi R^+-0} = \varphi\big|_{r^+=\xi R^++0}, \qquad \left.\frac{\partial\varphi}{\partial r^+}\right|_{r^+=\xi R^+-0} = \frac{k_{eff}}{k_f}\left.\frac{\partial\varphi}{\partial r^+}\right|_{r^+=\xi R^++0} \tag{43}$$

In addition, for the uniform wall heat flux case we required that in the center of the duct, $r^+ = 0$:

$$\frac{\partial\theta}{\partial r^+} = 0 \tag{44}$$

Also, for the uniform wall temperature case we required that at $r^+ = 0$:

$$\frac{\partial\varphi}{\partial r^+} = 0 \tag{45}$$

## RESULTS AND DISCUSSION

We discretized the governing equations by using a finite-difference method. The obtained linear equations with tridiagonal matrices were solved by the Gauss-Siedel iterations. We used relaxation to improve convergence. We used the following parameter values: $c_F = 0.55$, $C = 0.1$, $k_m / k_f = 1$, $Pr = 1$, $Pr_t = 1$, $R^+ = 10^3$, $\beta = 0$, $\mu_{eff} / \mu_f = 1$, $\xi = 0.5$, $\varphi = 0.95$.

In Fig. 2 we show velocity distributions in the duct computed for $\mathrm{Da} = 10^{-1}$ and $\mathrm{Da} = 10^{-3}$. We present results for rough or hydraulically smooth interface. For a large Darcy number ($10^{-1}$) the velocity in the clear fluid region for the case of rough interface is significantly smaller than for the smooth interface. This is explained by larger eddy viscosity in the clear fluid region for the case of a rough interface. For a smaller Darcy number ($10^{-3}$) the difference between the velocity profiles for the situations with rough and smooth interface is much smaller. This is because for a smaller Darcy number the interface is less rough (the equivalent sand-grain roughness of the interface is proportional to the square root of the Darcy number, see Eq. (18)).

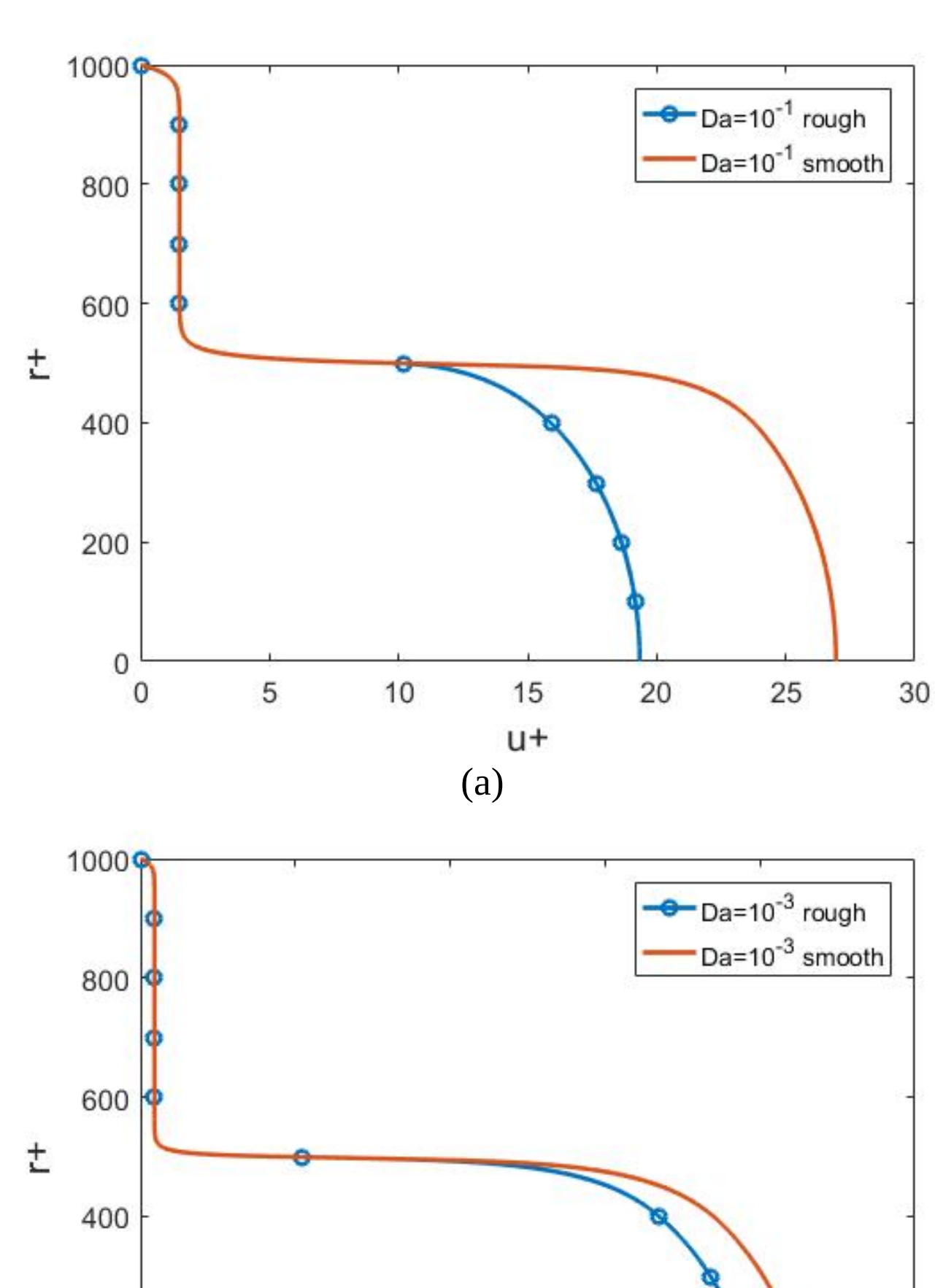

**Figure 2** Distributions of the dimensionless velocity for various Darcy numbers, for the situations with smooth and rough interface. (a) $\mathrm{Da} = 10^{-1}$, (b) $\mathrm{Da} = 10^{-3}$.

In Fig. 3 we displayed the dimensionless temperature distributions for both uniform wall heat flux and uniform wall temperature cases. We used $\mathrm{Da} = 10^{-1}$ and $\mathrm{Da} = 10^{-3}$ and presented the results for rough or hydraulically smooth interface. The results for the uniform wall heat flux and temperature are significantly different because the dimensionless temperature for these two cases is defined differently, by Eqs. (26) and (30), respectively.

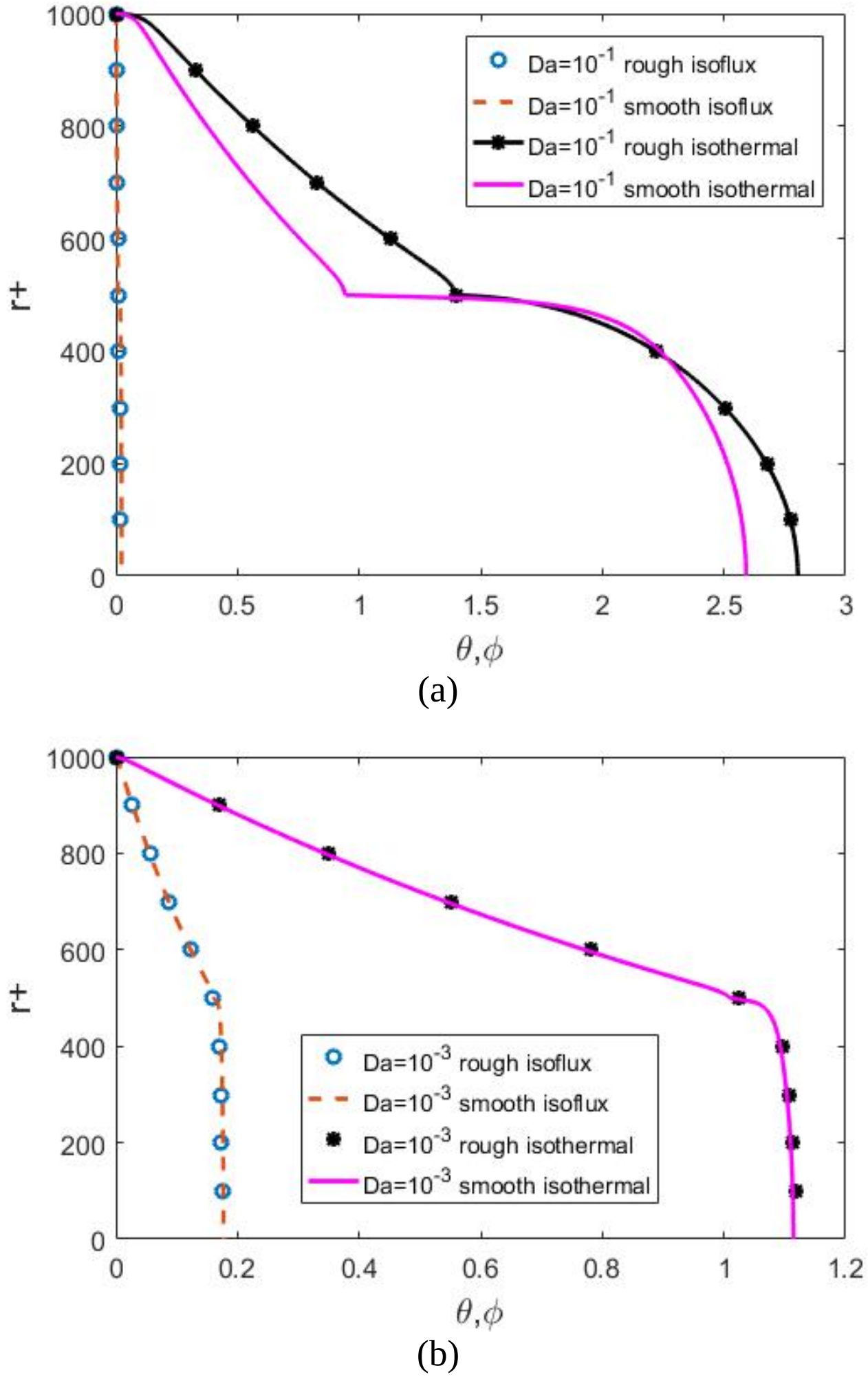

**Figure 3** Distributions of the dimensionless temperature for the situations with a constant wall heat flux and a constant wall temperature, for smooth and rough interface, for various Darcy numbers. (a) $\mathrm{Da} = 10^{-1}$, (b) $\mathrm{Da} = 10^{-3}$.

In Fig. 4 we show the effect of the Darcy number on the Nusselt number. Both situations (with a uniform wall heat flux and uniform wall temperature) are simulated. The rough interface leads to a stronger turbulence and larger Nusselt number. The difference between Nusselt number values for the rough and smooth interface cases becomes larger for larger values of the Darcy number. This is explained by the increase of the equivalent sand-grain roughness of the interface with the increase of the Darcy number.

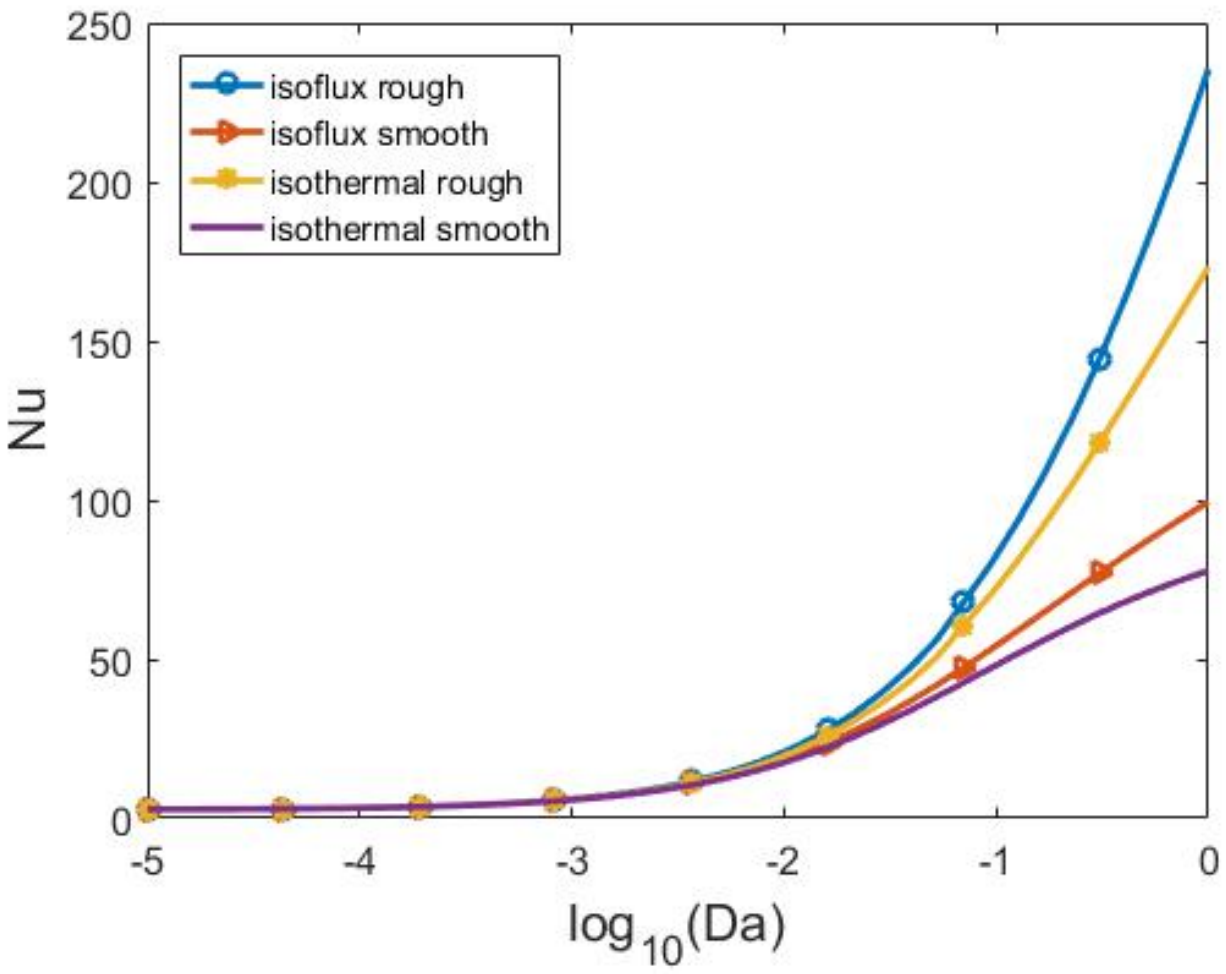

**Figure 4** The effect of the Darcy number on the Nusselt number for the situations with a constant wall heat flux and a constant wall temperature, for smooth and rough interface.

## CONCLUSIONS

We reviewed approaches of modeling turbulence in composite porous/fluid domains. Based on our recently obtained DNS results, due to the limitation of the size of turbulent eddies by the size of the pore, it is possible to model momentum transport in the porous region by the Brinkman-Forchheimer equation. The problem of modeling convection flow in a porous/fluid domain is then reduced to matching solutions of the Brinkman-Forchheimer equation in the porous region to the solution of an appropriate turbulence model at the interface between porous and fluid regions.

## ACKNOWLEDGMENT

The author acknowledges with gratitude the support of the National Science Foundation (award CBET-1642262) and the Alexander von Humboldt Foundation through the Humboldt Research Award.